\documentclass[doublecol]{epl2} 

\usepackage{graphicx}
\usepackage{color}
\usepackage{epsfig}
\usepackage{amssymb}
\usepackage{amsfonts}
\usepackage{amsmath}

\title{Proportional electroluminescence in two-phase argon and its relevance to rare-event experiments}

\author{A. Bondar\inst{1,2} \and A. Buzulutskov\inst{1,2}\footnote{Corresponding author. Email: A.F.Buzulutskov@inp.nsk.su} \and A. Dolgov\inst{2} \and V. Nosov\inst{1,2} \and L. Shekhtman\inst{1,2} \and E. Shemyakina\inst{1,2} \and A. Sokolov\inst{1.2}} \shortauthor{A. Bondar \etal}

\institute{
  \inst{1} Budker Institute of Nuclear Physics SB RAS, Novosibirsk, 630090, Russia\\
  \inst{2} Novosibirsk State University, Novosibirsk, 630090, Russia
}

\pacs{95.55.Vj}{Neutrino, muon,pion, and other elementary particle detectors; cosmic ray detectors} \pacs{61.25.Bi}{Liquid noble gases} \pacs{95.35.+d}{Dark matter}

\abstract{Proportional electroluminescence (EL) in gaseous Ar has for the first time been systematically studied in the two-phase mode, at 87 K and 1.00 atm. Liquid argon had a minor (56 ppm) admixture of nitrogen, which allowed to understand, inter alia, the effect of N$_2$ doping on the EL mechanism in rare-event experiments using two-phase Ar detectors. The measurements were performed in a two-phase Cryogenic Avalanche Detector (CRAD) with EL gap located directly above the liquid-gas interface. The EL gap was optically read out in the Vacuum Ultraviolet (VUV), near 128 nm (Ar excimer emission), and in the near Ultraviolet (UV), at 300-450 nm (N$_2$  Second Positive System emission), via cryogenic PMTs and a Geiger-mode APD (GAPD). Proportional electroluminescence was measured to have an amplification parameter of 109$\pm$10 photons per drifting electron per kV overall in the VUV and UV, of which 51$\pm$6\% were emitted in the UV. The measured EL threshold, at an electric field of 3.7$\pm$0.2 kV/cm, was in accordance with that predicted by the theory. The latter result is particularly relevant to DarkSide and SCENE dark matter search-related experiments, where the operation electric field was thereby on the verge of appearance of the S2 (ionization-induced) signal. The results obtained pave the way to the development of N$_2$-doped two-phase Ar detectors with enhanced sensitivity to the S2 signal.}

\begin{document}

\maketitle

\section{Introduction}

In two-phase detectors for dark matter search \cite{Zeplin3,Xenon10S2,Xenon100,Lux,Panda,Warp,Darkside} and coherent neutrino-nucleus scattering (CNNS) \cite{CoNu1,CoNu2} experiments, the S2 signal (induced by the primary ionization in the noble-gas liquid),  is detected through the effect of proportional electroluminescence (or proportional scintillations or else secondary scintillation) in the noble gas \cite{Rev1}. In low-mass ($<$10 GeV) dark matter search experiments and in particular in CNNS experiments, the significance of the ionization signal substantially increases, since it might remain the only one to be detectable \cite{Xenon10S2,CoNu1,CoNu2}. Accordingly, the problem of the effective detection of the electroluminescence signal (S2) becomes of paramount importance to low-threshold rare-event experiments.

Proportional electroluminescence in two-phase Ar detectors is produced when the primary ionization electrons, emitted from the liquid, drift in a moderate electric field in an electroluminescence (EL) gap located directly above the liquid-gas interface. In proportional electroluminescence the photon yield is basically proportional to the electric field, since the energy provided to the electrons by the electric field is almost fully expended in atomic excitations producing $Ar^{\ast}(3p^54s^1)$ and $Ar^{\ast}(3p^54p^1)$ states (the first one being metastable). These are followed by the photon emission in the Vacuum Ultraviolet (VUV), around 128 nm, due to excimer productions in three-body collisions and their subsequent decays \cite{PropELMech,PropELGAr,PropELSim}

\begin{equation}
\label{eq.1}
Ar^{\ast}(3p^54s^1)+2Ar\rightarrow Ar_2^{\ast}+Ar , \hspace{1ex} Ar_2^{\ast}\rightarrow 2Ar + h\nu ,
\end{equation}
and by the photon emission in the Near Infrared (NIR), at 690-850 nm, due to $Ar^{\ast}(3p^54p^1)\rightarrow Ar^{\ast}(3p^54s^1)+h\nu$ atomic transitions \cite{NirYield1,NirCRAD,NirYieldSim}. The EL yield in the NIR is by an order of magnitude lower than that in the VUV, due to the higher excitation threshold \cite{NirYield1,NirYieldSim}.

In presence of N$_2$ admixture in gaseous Ar at high ($\sim$1 atm) pressures the mechanism of proportional electroluminescence may change its nature, namely the excimer production (and hence the VUV emission) can be taken over by that of excited N$_2$ molecules in two-body collisions \cite{PropELArN2,ArN2Kin}, followed by their de-excitations through the emission of the so-called Second Positive System (SPS) \cite{ArN2SPS,N2SPSAir}, at 300-450 nm \cite{PropELArN2}:

\begin{eqnarray}
\label{eq.2}
Ar^{\ast}(3p^54s^1)+N_2\rightarrow Ar+N_2^{\ast}(C^3\Pi_u) ,
\\ N_2^{\ast}(C^3\Pi_u)\rightarrow N_2^{\ast}(B^3\Pi_g)+h\nu \hspace{1ex}. \nonumber
\end{eqnarray}
For simplicity in the following, the N$_2$ SPS emission in the near ultraviolet (UV) and visible spectral range is referred to as that in the UV. Note that the total amount of photons emitted both in reactions (1) and (2) remains constant, since one atomic excitation gives rise to one photon, emitted either in the VUV or UV (with a negligible addition in the NIR).

It is surprising that while proportional electroluminescence in gaseous Ar was rigorously studied at room temperature \cite{PropELGAr,PropELSim},  little is known about that in the two-phase mode. Moreover, the available data are rather scarce and confusing. Indeed, the EL yield in the VUV in two-phase Ar obtained in the WARP experiment and presented in  \cite{PropELGAr}, was reported to be an order of magnitude lower than the yield at room temperature. We will show in Section 3 that such data cannot be true.

Similarly, the EL yield in the NIR in two-phase Ar (at 87 K) was reported to be reduced, by an order of magnitude, compared to that at 163 K \cite{NirCRAD}. A plausible explanation of this effect will be given in Section 3.

Another motivation for this study comes from the recent results of the DarkSide dark matter search experiment \cite{Darkside}  and those of the energy calibration of two-phase Ar detectors \cite{LArIonYieldBern1,LArIonYieldScene}. In particular we will show in the following that the electric field values used in the EL gap in DarkSide \cite{Darkside} and SCENE \cite{LArIonYieldScene} experiments are dangerously close to the EL threshold determined in the present study, making it difficult to properly detect the S2 signal.

In this work we systematically study proportional electroluminescence in Ar in the two-phase mode, with a minor (56 ppm) admixture of N$_2$ that might be typical for large-scale liquid Ar experiments \cite{LargeLArRef}. The EL threshold and the electric field dependence of the absolute EL yield are measured. We show that the N$_2$ SPS emission provides a major part of the EL signal in these conditions.

The present study was performed in the course of the development of two-phase Cryogenic Avalanche Detectors (CRADs) of ultimate sensitivity for rare-event experiments \cite{RevCRAD} and their energy calibration \cite{LArIonYieldCRAD}. An advanced two-phase CRAD prototype was used in the measurements, with EL gap optically read out using cryogenic PMTs, and with combined THGEM/GAPD-matrix  multiplier  \cite{NirCRAD,RevCRAD,CRADMatrix} (THGEM is Thick Gas Electron Multiplier \cite{THGEMRev}; GAPD is Geiger-mode APD). Such a combined charge/optical readout of two-phase detectors would result in a higher overall gain at superior spatial resolution. It should be remarked that a novel design of the EL gap in two-phase detectors, when PMTs are located on the perimeter of the gap \cite{NirCRAD}, has been applied in this work. Accordingly, the verification of the performance of such an advanced two-phase CRAD was another motivation for this study.

\section{Experimental setup and measurement procedures}

\begin{figure}
\includegraphics[width=0.99\columnwidth]{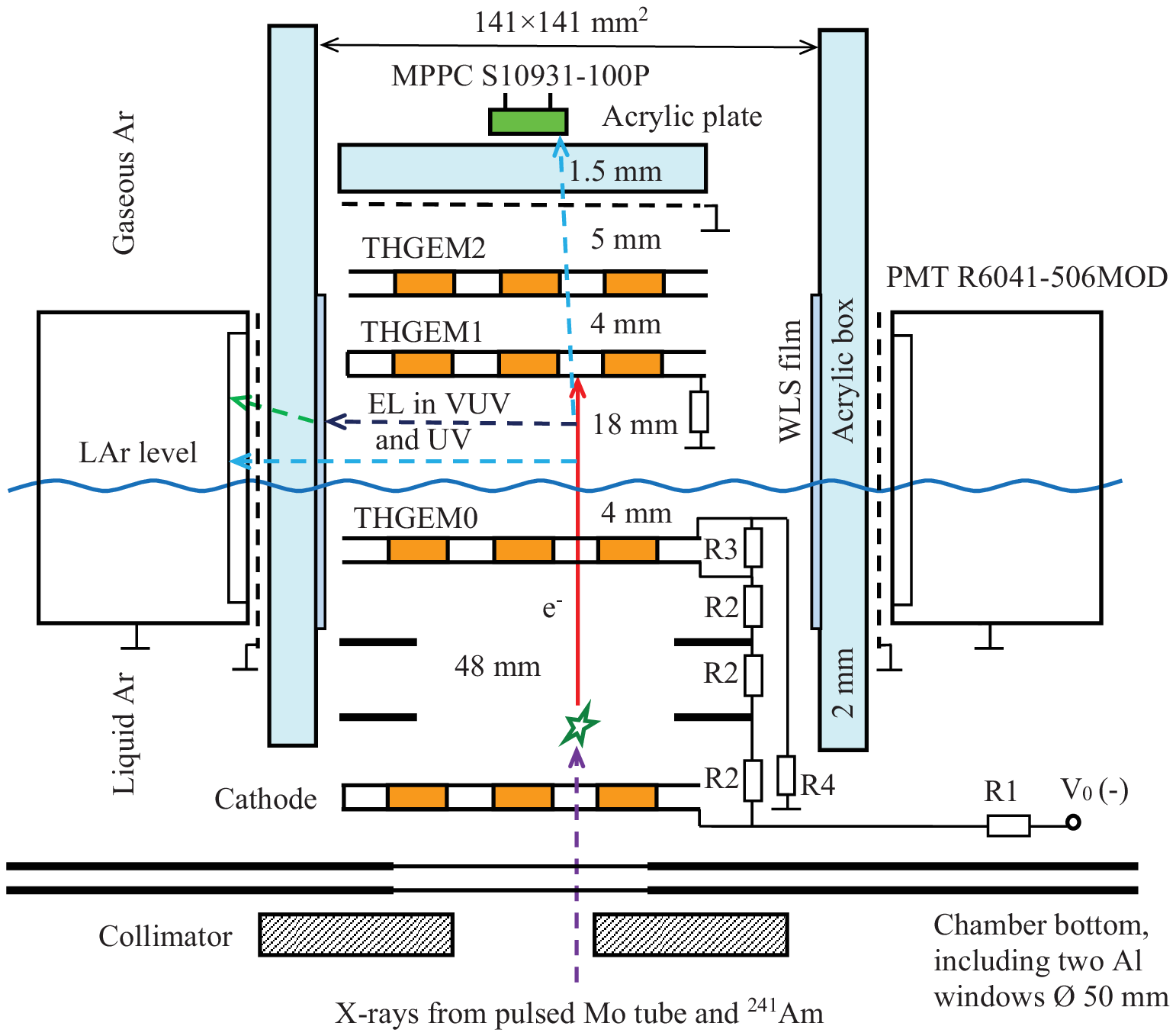}
\caption{Schematic view of the experimental setup (not to scale). Only the central MPPC of the 5$\times$5 matrix is shown. The resistors of the voltage divider have the following values: R1, R2, R3 and R4 is 80, 40, 4 and 600 MOhm respectively.}
\label{Setup}
\end{figure}

Fig.~\ref{Setup} shows the experimental setup, representing the first working prototype of the two-phase CRAD with both EL gap and THGEM/GAPD-matrix multiplier. It comprised a vacuum-insulated 9 l cryogenic chamber filled with 2.5 liters of liquid Ar. Ar was taken from a bottle with a specified purity of 99.998\% (specified N$_2$ content $<$10 ppm). During each cooling procedure Ar was purified from electronegative impurities by Oxisorb filter, providing electron life-time in the liquid $>$70 $\mu$s. The detector was operated in two-phase mode in the equilibrium state, at a saturated vapor pressure of 1.000$\pm$0.003 atm and at a temperature of 87.3 K.

After each cryogenic measurement, the Ar was liquified from the chamber back to a stainless steel bottle cooled with liquid nitrogen, so that the N$_2$ content remained constant throughout the entire measurement campaign. The latter lasted 5 months during which the setup (operated on a closed loop) was repeatedly evacuated but neither baked nor purified from N$_2$, resulting in a certain N$_2$ content established due to combined effect of the residual gas and internal outgassing in the bottle and cryogenic chamber.  At the end of the campaign, the bottle with Ar was connected to a baked high-vacuum (10$^{-9}$ mbar) system equipped with a Residual Gas Analyzer Pfeiffer-Vacuum QME220, where the N$_2$ content in Ar was measured in a flow mode at a pressure of $10^{-4}$ mbar: it amounted to 56$\pm$5 ppm. In the two-phase mode at 87 K, this value corresponds to the N$_2$ content of 56 ppm in the liquid and 151 ppm in the gas phase, according to "Raoult" law \cite{TPArN2CRAD}.

The cryogenic chamber included a cathode electrode (made from a THGEM plate), two field-shaping electrodes and a THGEM0, immersed in a 55 mm thick liquid Ar layer. These 4 elements were biased through a resistive high-voltage divider placed within the liquid, forming a drift region in liquid Ar, 48 mm long. A 4 mm thick liquid Ar layer above the THGEM0 acted as an electron emission region. A double-THGEM assembly, consisting of a THGEM1 and THGEM2, was placed in the gas phase above the liquid. The EL gap (the EL region), 18 mm thick, was formed by the liquid surface and the THGEM1 plate; the latter was grounded through a resistor acting as an anode of the gap. The rigidity of the THGEM0 and THGEM1 plates provided the good flatness of the EL gap even under high-field conditions. All electrodes had the same active area, of 10$\times$10 cm$^2$.

The voltage applied to the divider varied from 11 to 22 kV, producing an electric drift field in liquid Ar of 0.34-0.68 kV/cm, electric emission field of 2.6-5.1 kV/cm and electric field in the EL gap of 4.0-8.0 kV/cm. Using such a voltage divider, the THGEM0 was biased in a way to provide the effective transmission of drifting electrons from the drift region to that of electron emission: the electrons drifted successively from a lower to higher electric field region, with a field ratio of about 3 at each step, which in principle should result in electron transmittance through the THGEM0 approaching 100\% \cite{ETrans}. In addition, the high emission field, exceeding 2 kV/cm, guaranteed a full electron extraction from the liquid into the gas phase \cite{Rev1}. The average drift time of the electrons across the drift, emission and EL regions varied from about 25 to 35 $\mu$s, depending on the applied electric fields.

There were two ways of optical readout of the EL gap. Firstly, the gap was viewed by four compact cryogenic 2-inch PMTs R6041-506MOD \cite{Hamamatsu}, located on the perimeter of the gap at a distance of 77 mm from its center. To prevent discharges and field penetration from the high-voltage region of the EL gap, the PMTs were electrically insulated from the gap by a grounded mesh and an acrylic protective box of a rectangular shape. The box had an inner size of 141$\times$141 mm$^2$ and wall thickness of 2 mm. To convert the VUV into the blue light, four wavelength shifter (WLS) films based on TPB (tetraphenyl-butadiene) were deposited on the inner box surface facing the EL gap, in front of each PMT. The WLS films were deposited by spreading the TPB and polystyrene mixture (in 1:3 proportion) in toluene solution on the box walls, allowing the toluene to evaporate \cite{TPB1,TPB2}; the effective TPB thickness in the film was 0.26 mg/cm$^2$. Each PMT had a bialkali photocathode of 45 mm diameter. The PMTs were powered through voltage dividers mounted on their bases.

Secondly, the scintillation in the spectral range other than the VUV, i.e. in the near UV (and visible and NIR range, if any) could be recorded using a 5$\times$5 matrix of GAPDs of the MPPC (Multi-Pixel Photon Counter) S10931-100P type \cite{Hamamatsu,CryoGAPD}, placed in the gas phase behind the THGEM2. The MPPCs were electrically insulated from the latter by a grounded grid and an acrylic plate, 1.5 mm thick. Each MPPC had a 3$\times$3 mm$^2$ active area. In this paper, the double-THGEM multiplier was not operated in a charge multiplication mode and the MPPCs were not involved in optical readout of the avalanche scintillations; this will be the subject of the forthcoming studies. Instead, only the central MPPC of the matrix was used to directly record proportional electroluminescence from the EL gap through the double-THGEM assembly, the latter acting as a shadow mask.

The detector was irradiated from outside through a collimator and two aluminium windows (each 1 mm thick) by X-rays from a pulsed X-ray tube with Mo anode operated at a voltage of 40 kV (at a rate of 240 Hz) \cite{XRayYield}. The pulsed X-rays were attenuated by X-ray filters resulting in the incident X-ray energy of 30-40 keV. The X-ray pulse was strong enough, to provide a measurable ionization charge in the EL gap (tens of thousands electrons), and sufficiently fast (0.5 $\mu$s), to provide a reasonable time resolution.

Three types of signals were recorded: the optical signal from the PMTs, the charge signal from the THGEM1 and the optical signal from the MPPC. The optical signal from the four PMTs, called the total PMT signal, was obtained  as a linear sum of all the PMT signals (using CAEN N625 unit), amplified with a linear amplifier with a shaping time of 200 ns. The THGEM1 charge signal was recorded using a charge-sensitive preamplifier followed by a shaping amplifier with a time constant of 1 $\mu$s. The MPPC optical signal was recorded using a fast amplifier with a shaping time of 40 ns. The DAQ system included a 4-channel oscilloscope LeCroy WR HRO 66Zi. In the case of pulsed X-rays, the trigger was external, provided by a pulsed X-ray tube generator.

The amplitude of each of the three signal types was measured by time-integration of the appropriate part of the pulse. The amplitude of the PMT signal was expressed in the number of photoelectrons (p.e. or pe) recorded on the PMT photocathodes ($N_{pe}$), by normalizing to the average single p.e. amplitude, obtained from the single p.e. amplitude spectra averaged over all the PMTs. The single p.e. spectra were provided by the PMT noise signals. The amplitude of the charge (THGEM1) signal was expressed in the number of electrons ($N_e$), using the amplifier circuit calibration including a charge injection with the help of a pulse generator and a precise capacitance. The amplitude of the MPPC signal was expressed in the number of primary photoelectrons, by normalizing to the average amplitude of the MPPC "single p.e." spectrum. It should be remarked that there was a substantial ($\sim$30\%) crosstalk between the MPPC pixels, appearing as the second and third peaks in the single p.e. spectrum; its contribution was taken into account.

\begin{figure}
\includegraphics[width=0.99\columnwidth]{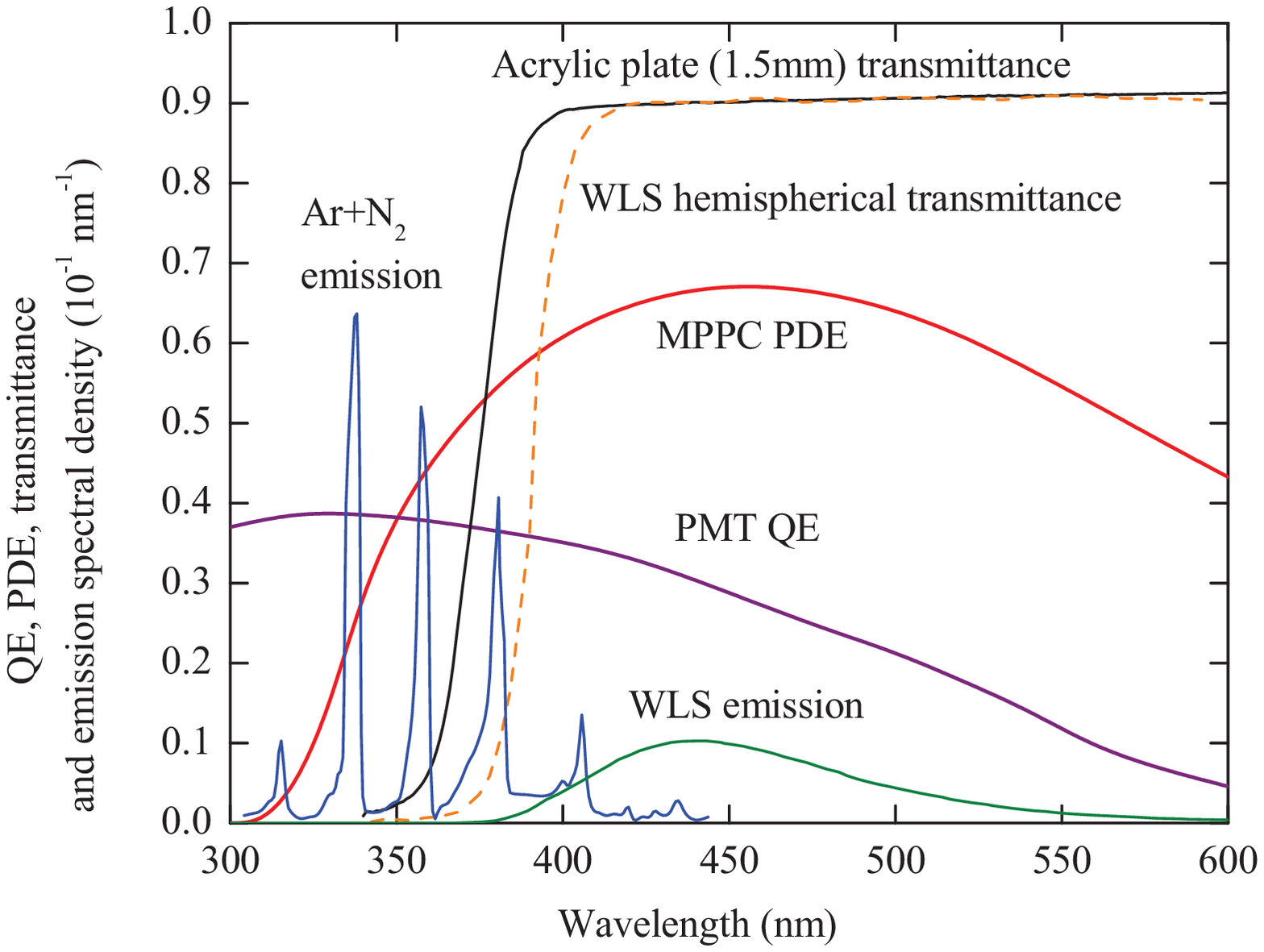}
\caption{Quantum efficiency (QE) of the PMT R6041-506MOD at 87 K obtained from \cite{CryoPMT1,Hamamatsu} using a temperature dependence derived there, Photon Detection Efficiency (PDE) of the MPPC S10931-100P \cite{CryoGAPD,Hamamatsu}, transmittance of the 1.5 mm thick acrylic plate in front of the MPPC measured by us and hemispherical transmittance of the WLS \cite{TPB2}.  Also shown are the EL emission spectrum of Ar doped with N$_2$ (0.2\%) in the near UV and visible range measured at room temperature and high ($\sim$1 atm) pressure \cite{PropELArN2} and that of the WLS (TPB in polystyrene) \cite{TPB1}.} \label{Spectrum}
\end{figure}

To convert the measured amplitudes into the number of photons emitted in the VUV and UV, the light collection efficiency ($\varepsilon$) should be determined. For this procedure, Fig.~\ref{Spectrum} presents optical spectra of the PMT Quantum Efficiency (QE) \cite{CryoPMT1}, MPPC Photon Detection Efficiency (PDE) \cite{CryoGAPD}, acrylic plate transmittance (measured by us) and WLS (TPB in polystyrene) hemispherical transmittance \cite{TPB2}. In addition, the EL emission spectrum of Ar doped with N$_2$ \cite{PropELArN2} and that of the WLS (TPB in polystyrene) \cite{TPB1} are presented.

\section{Results and discussion}

The sensitivity of a particular detector is characterized by the EL gap yield. In our case it can be defined as the number of photoelectrons recorded by the PMTs or the MPPC ($N_{pe}$), per drifting electron in the EL gap: $Y_{EL gap}=N_{pe}/N_e $. Here $N_e$ is the charge expressed in the number of electrons, drifting in the EL gap and producing proportional electroluminescence.

\begin{figure}
\includegraphics[width=0.99\columnwidth]{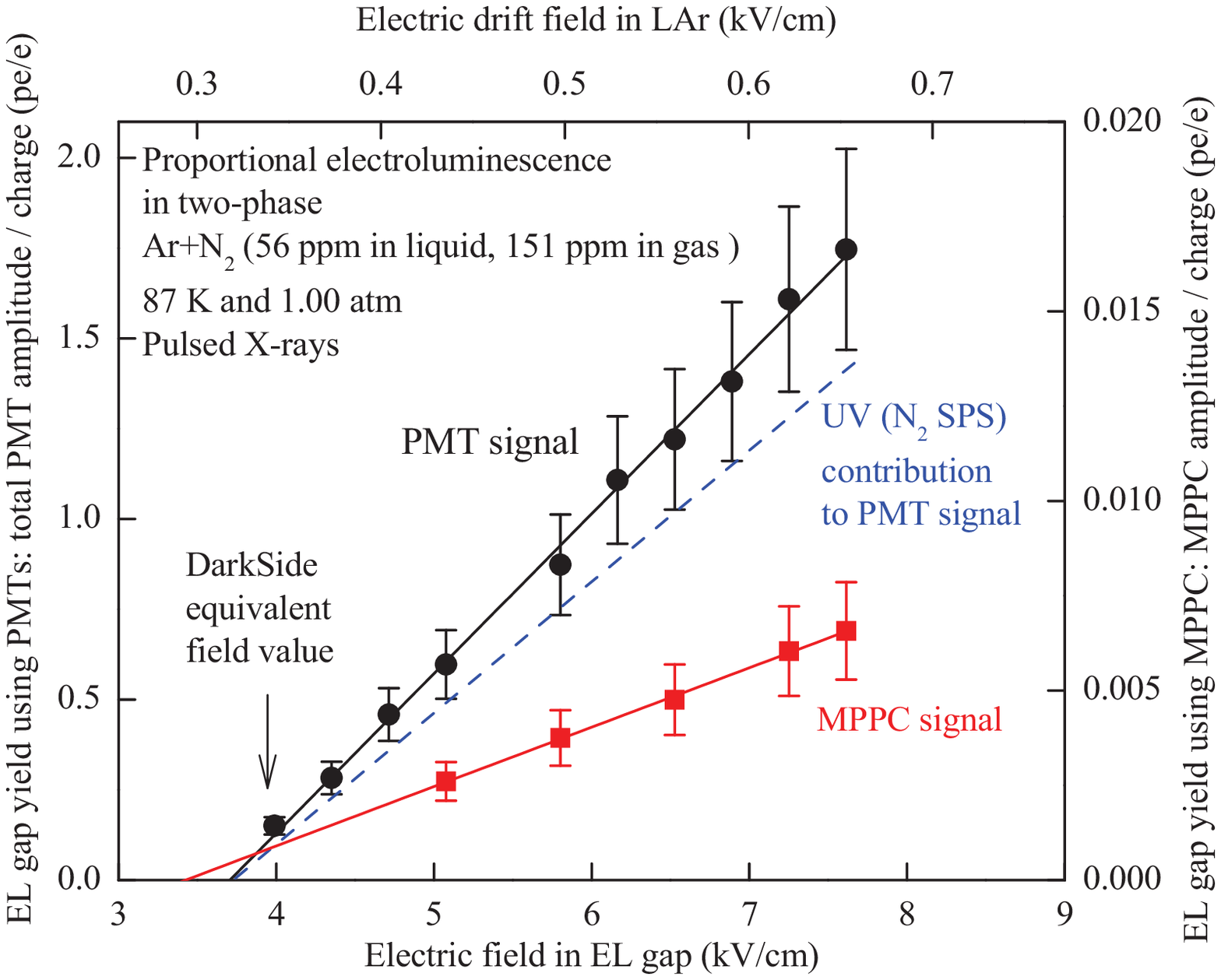}
\caption{EL gap yield measured using PMT signal (left scale) and that of MPPC (right scale) as a function of the electric field in the EL gap. The arrow indicates the electric field value equivalent to that of DarkSide experiment \cite{Darkside}, i.e. reduced to the gas density of the present study. The electric drift field in liquid Ar is also shown on the top axis. The dashed line indicates the contribution of proportional electroluminescence in the UV (i.e. due to the N$_2$ SPS emission) to the PMT signal, deduced from the EL yield using MPPC signal.}
\label{ELGapYield}
\end{figure}

Fig.~\ref{ELGapYield} shows the EL gap yield as a function of the electric field in the gap, measured using PMT and MPPC signals. We can clearly see a linear dependence of the yield on the electric field, that confirms its proportional electroluminescence nature. One can also see that proportional electroluminescence measured with the PMTs had a threshold at an electric field of 3.7$\pm$0.2 kV/cm. Within the experimental uncertainties, the same threshold had proportional electroluminescence measured with the MPPC, clearly indicating that the PMT and MPPC signals are linked.

An important conclusion is that the electric field value in the EL gap, equivalent to that of DarkSide dark matter search experiment \cite{Darkside} (reduced here to the gas density of the present study), of 3.9 kV/cm, is dangerously close to the EL threshold. This is also true for the SCENE energy calibration experiment \cite{LArIonYieldScene}, operated at a similar field. Such an operation, on the verge of the signal appearance, may result in improper detection of the S2 signal in those experiments. In particular we assume that the strong radial dependence of the S2 amplitude observed in DarkSide experiment \cite{Darkside} might be due to the sensitivity of the S2 signal, enhanced at the EL threshold, to even minor variation of the electric field.

The presence of the MPPC signal leads us to a conclusion that there was an additional contribution of the photon emission in the spectral range other than that of the VUV. Indeed, the calculations showed that proportional electroluminescence in the VUV itself cannot explain the EL gap yield measured using the PMTs: we lack a factor of 2.7 in the yield value. This additional contribution cannot be due to the emission in the NIR, since the latter was too weak (at higher temperatures) and, in addition, it was found to be further suppressed at 87 K \cite{NirCRAD}. On the other hand as mentioned in the Introduction, even minor admixture of N$_2$ to Ar might give rise to the EL emission in the near UV where the MPPC has high sensitivity (Fig.~\ref{Spectrum}). Accordingly, we adopt here the hypothesis that there were two principal sources of the EL emission in two-phase Ar in our experiment: first, due to the Ar excimer emission in the VUV and, second, due to the N$_2$ SPS emission in the UV (see reactions (\ref{eq.1}) and (\ref{eq.2})).

This hypothesis requires three spectral components contributing to the EL gap yield using the PMTs (see Fig.~\ref{Spectrum}), namely those of the VUV and UV, absorbed and re-emitted in the WLS, and that of the UV escaping absorption in the WLS. Fortunately, the EL yield due to the emission in the UV was directly measured by the MPPC. Hence one can calculate the contribution of the two UV components to the PMT signal using the absolute EL yield and the light collection efficiency.

The absolute EL yield is defined as the number of emitted photons ($N_{ph}$) normalized to the number of drifting electrons producing electroluminescence and to the electron drift path ($d$): $ Y_{EL}=N_{ph}/N_e/d$. The number of photons recorded by the PMTs is defined as $N_{ph}=N_{pe}/\varepsilon/CE/QE$, for the VUV and UV components re-emitted by the WLS, and as  $N_{ph}=N_{pe}/\varepsilon/QE$ for the UV component recorded directly.  The number of photons recorded by the MPPC is defined as $N_{ph}=N_{pe}/\varepsilon/PDE$. Here $\varepsilon$ and $CE$ are the light collection and WLS conversion efficiencies, for a given emission component; $QE$ and $PDE$ are the PMT QE and MPPC PDE averaged over the WLS and N$_2$ SPS emission spectrum respectively and convoluted, if needed, with the acrylic plate or WLS hemispherical transmittance spectrum (Fig.~\ref{Spectrum}).

To determine the light collection efficiencies, four Monte-Carlo simulation procedures were performed. The first and the second ones simulated the emission of the photons in the EL gap of either the VUV (128 nm) or the UV component absorbed in the WLS, the latter amounting to 75\% of the N$_2$ SPS, at 300-400 nm, obtained from convolution of the N$_2$ SPS and WLS hemispherical transmittance spectra of Fig.~\ref{Spectrum}. The propagation of photons to the WLS film, their conversion by the WLS film to the visible light photons and their further propagation to the PMT photocathodes were simulated. The light collection efficiency amounted to 3.1$\times10^{-3}$ for the VUV and 2.5$\times10^{-3}$ for the UV component respectively. Here we used $CE=0.58$  at 128 nm \cite{TPB1} and $CE=0.40$ at 300-400 nm \cite{TPB1,TPB2,TPB3}.


The third procedure simulated the emission of the photons of the UV component escaping absorption in the WLS (amounting to 15\% of the N$_2$ SPS, at 400-450 nm) and thus directly propagating to the PMTs. The light collection efficiency for this component, of 5.4$\times10^{-2}$,  turned out to be considerably higher than that of the light re-emitted by the WLS, by about a factor of 20, due to the absence of re-emission and total internal reflection losses.

The fourth procedure simulated the emission of the photons of the UV component and their propagation to the central MPPC through the two THGEMs 
(amounting to 31\% of the N$_2$ SPS, at 360-450 nm). In this case the light collection efficiency amounted to 8.6$\times10^{-5}$.

\begin{figure}
\includegraphics[width=0.99\columnwidth]{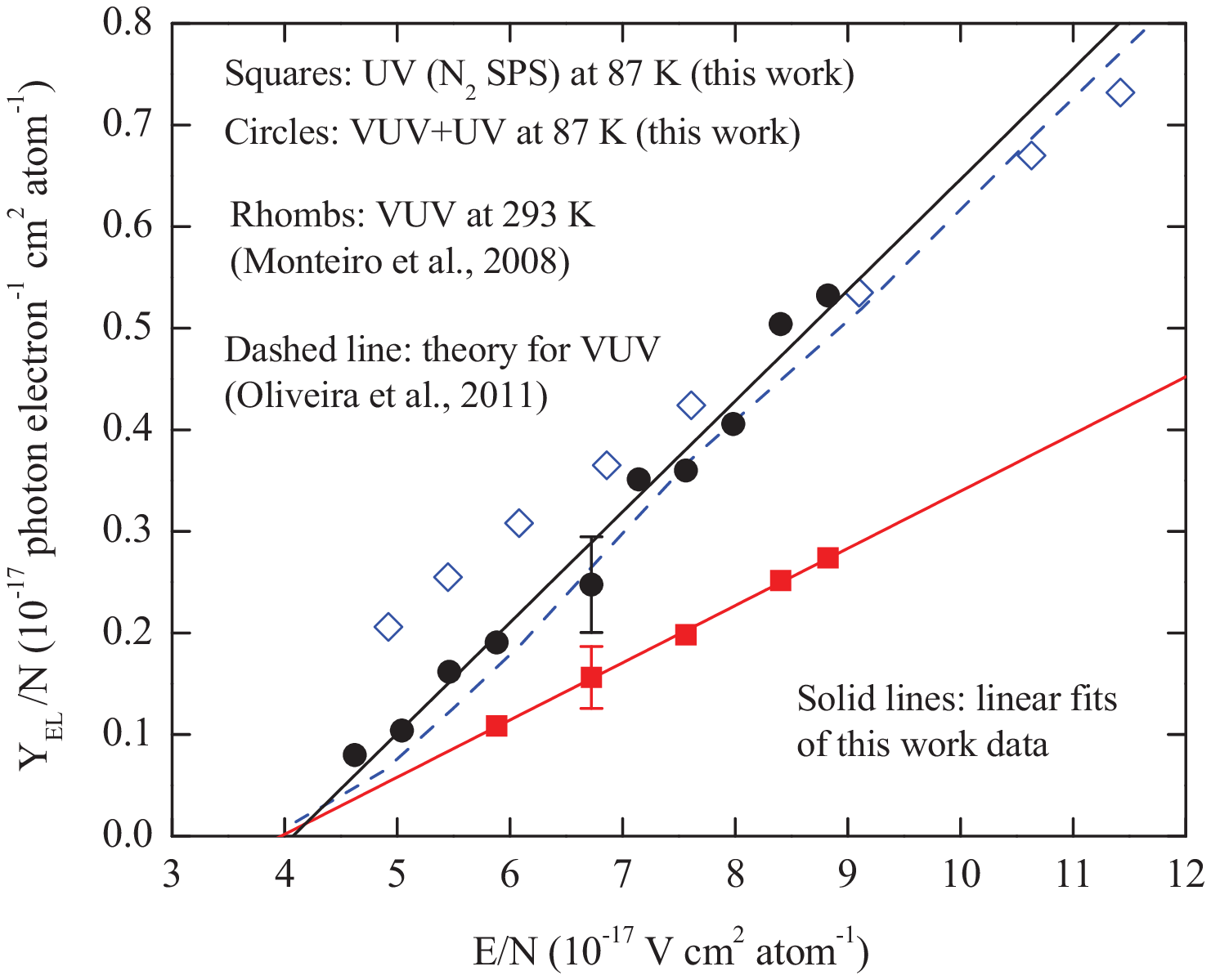}
\caption{Reduced electroluminescence yield as a function of the reduced electric field in the UV (N$_2$ SPS emission) and overall in the VUV (Ar emission) and UV, determined in this work in gaseous Ar doped with N$_2$ (56 ppm in the liquid and 151 ppm in the gas phase), in the two-phase mode at a temperature of 87 K and pressure of 1.00 atm. For comparison, the yields in gaseous Ar in the VUV obtained experimentally at 273 K (Monteiro et al. \cite{PropELGAr}) and theoretically (Oliveira et al. \cite{PropELSim}) are presented. The solid lines are the linear fits to the data points of the present study.}
\label{ELYield}
\end{figure}

To compare our results to those measured at different temperatures and pressures, the reduced EL yield ($Y_{EL}/N$) is plotted in Fig.~\ref{ELYield} as a function of the reduced electric field ($E/N$), where both quantities are normalized to the atomic concentration ($N$). Here $E/N$ is given in Td [10$^{-17}$ V cm$^2$ atom$^{-1}$] and $Y_{EL}/N$ in [10$^{-17}$ photon electron$^{-1}$ cm$^2$ atom$^{-1}$].

The linear dependence of the EL yield in the UV (N$_2$ SPS) on the electric field, derived from the MPPC data, is described by the following equation (Fig.~\ref{ELYield}):

\begin{equation}
\label{eq.3}
Y_{EL}/N = 0.056E/N - 0.223 .
\end{equation}
The slope of the line defines the amplification parameter of proportional electroluminescence. Accordingly in the UV, it amounts to 56$\pm$7 photons per kV. Using this parameter, we can calculate the desired contribution of the two UV components to the PMT signal: it provides the major 82\% of the signal (dashed line in Fig.~\ref{ELGapYield}), the rest of the signal being provided by emission in the VUV. The detailed analysis showed that such a domination is provided mostly by the direct UV component escaping absorption in the WLS, due to the considerably higher light collection efficiency as discussed above.

Once the VUV contribution is defined, one can determine the total EL yield: it is shown in Fig.~\ref{ELYield}. The linear dependence of the total EL yield on the electric field, overall in the VUV (Ar excimer emission) and UV (N$_2$ SPS emission), is described by the following equation:

\begin{equation}
\label{eq.4}
Y_{EL}/N = 0.109E/N - 0.444 .
\end{equation}
For comparison, the yields in gaseous Ar in the VUV obtained experimentally at 273 K \cite{PropELGAr} and theoretically \cite{PropELSim} are also shown.

We can state that the high EL yield measured here refutes the previous result of the WARP group presented in \cite{PropELGAr}. One can also conclude that both the EL threshold and the amplification parameter, of 109$\pm$10 photons per kV,  measured in this work at 87 K are in good agreement with the theory. This fact indicates that the total number of photons, emitted no matter in the VUV or UV, remains unchanged, with good accuracy being equal to the number of the excited Ar atoms as predicted by the theory \cite{PropELSim}. Such an amazing agreement with the theory was obtained thanks to fine-tuned balance between the VUV and UV emission components, that can hardly be obtained accidentally, thus strongly supporting the hypothesis of the combined Ar excimer and N$_2$ SPS emission.

Since at room temperature the N$_2$ emission in Ar was not observed at such a small N$_2$ content \cite{PropELArN2}, we may suppose that the effect of N$_2$ doping is enhanced at lower temperatures.

Now we can explain the effect of the EL yield reduction in the NIR observed in Ar at 87 and 126 K \cite{NirCRAD}. This reduction is supposed to be induced by a minor N$_2$ admixture due to the reaction similar to that of (\ref{eq.2}), when the excited $Ar^{\ast}(3p^54p^1)$ states providing emission in the NIR, transfer their excitation to the $N_2^{\ast}(C^3\Pi_u)$ states emitting in the UV. Because of the NIR filter, the GAPDs used there were not sensitive to the UV \cite{NirCRAD}. Together, these factors may explain why the EL yield in the NIR progressively decreased with decreasing temperature.


\section{Conclusion}

Proportional electroluminescence (EL) in gaseous Ar has for the first time been systematically studied in the two-phase mode, at 87 K and 1.00 atm. Ar had a minor admixture of N$_2$, of 56 ppm in the liquid and 151 ppm in the gas phase,  which allowed to understand, inter alia, the effect of N$_2$ doping on the EL mechanism. Such a N$_2$ content might be quite typical for large-scale liquid Ar experiments.

Proportional electroluminescence in two-phase Ar at 87 K was measured to have an amplification parameter of 109$\pm$10 photons per drifting electron per kV overall in the VUV and UV, of which 51$\pm$6\% were emitted in the UV (N$_2$ Second Positive System). This result indicates that the effect of N$_2$ doping on proportional electroluminescence in Ar is enhanced at lower temperatures.

In addition, the S2 signal response was found to be substantially enhanced due to the fraction of the N$_2$ emission spectrum recorded directly, i.e. escaping re-emission in the WLS film and thus having a considerably higher (by a factor of 20) light collection efficiency. This observation paves the way to the development of N$_2$-doped two-phase Ar detectors with enhanced sensitivity to the S2 signal.

The overall EL amplification parameter and the EL threshold (at an electric field of 3.7$\pm$0.2 kV/cm) measured in this work were in accordance with those predicted by the theory. The result on the EL threshold is particularly relevant to DarkSide \cite{Darkside} and SCENE \cite{LArIonYieldScene} dark matter search-related experiments, where the operation electric field was thereby on the verge of appearance of the S2 signal.


We thank A. Barnyakov, A. Chegodaev and R. Snopkov for technical support. This study was supported by Russian Science Foundation (project N 14-50-00080).

\end{document}